\documentclass{JHEP}
\usepackage{amsmath}
\usepackage{amssymb,amsfonts}
\usepackage{epsfig}

\newcommand{\nn}{\nonumber}
\renewcommand{\th}{\vartheta}

\newcommand{\Zint}{\mathbb{Z}}
\newcommand{\Real}{\mathbb{R}}

\title{(Anti-)Instantons and \\the Atiyah-Hitchin Manifold}

\author{Amihay Hanany\\
Center for Theoretical Physics,
\\ Massachusetts Institute of Technology\\ Cambridge, MA 02139, USA\\
E-mail: \email{hanany@mit.edu} }

\author{Boris Pioline\footnote{On
leave of absence from LPTHE, Universit{\'e} Pierre et Marie Curie,
PARIS VI and Universit{\'e} Denis Diderot, PARIS VII, Bo\^{\i}te
126, Tour 16, 1$^{\it er}$ {\'e}tage, 4 place Jussieu, F-75252
Paris CEDEX 05, FRANCE}\\
Jefferson Physical Laboratory, Harvard University\\
Cambridge, MA 02138, USA\\
E-mail: \email{pioline@physics.harvard.edu}}

\abstract{The Atiyah-Hitchin manifold arises in many different
contexts, ranging from its original occurrence as the moduli space
of two $SU(2)$ 't Hooft-Polyakov monopoles in 3+1 dimensions, to
supersymmetric backgrounds of string theory. In all these
settings, (super)symmetries require the metric to be hyperk\"ahler
and have an $SO(3)$ transitive isometry, which in the
four-dimensional case essentially selects out the Atiyah-Hitchin
manifold as the only such smooth manifold with the correct
topology at infinity. In this paper, we analyze the exponentially
small corrections to the asymptotic limit, and interpret them as
infinite series of instanton corrections in these various
settings. Unexpectedly, the relevant configurations turn out to be
bound states of $n$ instantons and $\bar n$ anti-instantons, with
$|n-\bar n|=0,1$ as required by charge conservation. We propose
that the semi-classical configurations relevant for the higher
monopole moduli space are Euclidean open branes stretched between
the monopoles.}

\preprint{hep-th/0005160\\HUTP-00/A15\\MIT-CTP-2983}

\begin{document}

\section{Introduction}
The advent of the now mundane dualities of supersymmetric field and string
theories has made it possible to obtain a wealth of
non-perturbatively exact results for various couplings in the
low-energy effective action of these theories, all of them
severely constrained by supersymmetry. In the seminal case of the
Coulomb phase of $N=2$ four-dimensional gauge theories
\cite{Seiberg:1994rs}, the holomorphicity of the prepotential
together with electric-magnetic duality was sufficient to fix the
dynamics of the vector-multiplets at two-derivative order for all
values of the (dimensionally transmuted) gauge coupling
$\Lambda^{b_0}=\mu^{b_0} \exp(-{8\pi^2}/{g^2_{YM}})$, and the
non-perturbative corrections of order $\Lambda^{n b_0}e^{i
n\theta}$ to the one-loop prepotential were identified as the
contribution from $n$ Yang-Mills instantons. In a similar manner,
exact higher derivative couplings in type I and type II string theories
have been obtained from the requirements of S-duality and
harmonicity (see \cite{Kiritsis:1999ss} for a review and 
references), and shown to
encapsulate the contributions from bound states of arbitrary
number $n$ of D-instantons, plus its complex-conjugate series of
anti-instanton contributions.

While holomorphy or harmonicity give powerful constraints on the
half-BPS--saturated couplings of vector-multiplets of $N=2$
supersymmetry (and presumably any multiplets of higher
supersymmetry), the couplings of hypermultiplets are by no means
as well understood, even though they occur with the same amount of
supersymmetry. The reason is that the hyperk\"ahler or
quaternionic constraints on the moduli space of these fields lack
a concise and tractable formulation (except perhaps for the
twistor methods, which are holomorphic with respect to an
auxiliary variable; see \cite{Hitchin:1987ea} for a review).
Indeed, there are to date no explicitly known non-homogeneous
quaternionic manifolds, and very few explicit examples of
hyperk\"ahler manifolds, all of them one-dimensional (in
quaternionic units) and with a large number of isometries. Among
them are the (multi) Eguchi-Hanson and Taub-NUT gravitational
instantons, with asymptotic geometry $\Real^4/\Zint_k$ and $\Real^3
\times S^1$ respectively, which both possess a triholomorphic
$U(1)$ isometry (times $SO(3)$ in the single instanton case; see
\cite{Bakas:1997gf,Bakas:2000wq} for a review). Hence, they can be obtained
from a harmonic function, which is a sum of a finite number of
pole terms in both cases (plus a constant for Taub-NUT).
By combining infinite series of poles, it is possible to generate
a non-perturbative behaviour, and indeed it was shown by Ooguri
and Vafa that such a space describes the quantum-corrected moduli
space of the conifold hypermultiplet \cite{Ooguri:1996me}. In the
weak coupling (or asymptotic) limit, 
one recovers a series of $n$ D-instanton effects,
together with its complex-conjugate series of anti-instanton
effects, as in the Type IIB case above.

The only explicit example of four-dimensional hyperk\"ahler space
without triholomorphic isometry (but with an $SO(3)$ group of
``rotational symmetries'' not rotating the three K\"ahler forms)
is the Atiyah-Hitchin manifold, first introduced in the context of
the moduli space of two BPS monopoles in a 3+1 $SU(2)$ gauge
theory with a Higgs field in the adjoint \cite{Atiyah:1985dv}.
This space is one-dimensional (after omitting a trivial
center-of-mass factor of $\Real^3\times S^1$), and consists of
three relative positions of the monopoles (measured in units of
the $W$-boson mass \footnote{We normalize the kinetic term
of the Higgs field to $(\nabla\phi)^2/g_{YM}^2$.} $M_W=\phi$)
and a relative phase $\sigma$. The
metric on that space controls the slow motion of the two monopoles
\cite{Manton:1982mp}, which primarily interact through the
exchange of massless photons and scalars at long distance
\cite{Gibbons:1986df}. In this regime, the moduli space reduces to
a Taub-NUT space, albeit with a negative mass parameter, and hence
a singularity at finite distance $r=2$ between the monopoles on
the order of the Higgs vev. The metric is independent of the gauge
coupling, there are therefore no quantum corrections to this
motion, whether perturbative or not. There are however corrections
to the long-range interaction due to the exchange of massive 
$W$-bosons and Higgs field, which dominates when they come close to
each other and resolves the singularity. The exact metric was
derived in \cite{Atiyah:1985dv} (see also \cite{Atiyah:1988jp}) on
the basis of $SO(3)$ isometry and self-duality (which expresses
hyperk\"ahlerity in 1 dimension), in terms of elliptic functions.
The deviation from the Taub-NUT limit is exponentially small in
the distance between the monopoles, and is most easily expressed
as a deviation to particular components of the Riemann tensor
which we shall make precise later,
\begin{eqnarray}
\label{1inst}
R_{(0)}&=&\frac{4}{(r-2)^3}+ \dots \\
R_{(+2)}&=&\frac{8(-3+9r-6r^2+r^3)}{(r-2)^2}
e^{-r+ i\sigma} + \dots \\
R_{(-2)}&=&\frac{8(-3+9r-6r^2+r^3)}{(r-2)^2}
e^{-r- i\sigma} + \dots
\end{eqnarray}
where the dots denote subleading corrections of order $e^{-2r}$.
The exponential terms in this expression can be interpreted as
the semi-classical effect of the Euclidean worldline of a massive
$W$-boson stretching between the two monopoles. In the following,
we will be mostly interested in the structure of the subleading
terms in this expansion, displayed in \eqref{Rq} below,
which will reveal the interplay between
instantons and anti-instantons.

While the above occurrence of the Atiyah-Hitchin manifold was purely
classical, the same manifold arises in many other instances
in string or field theory,
where the radial parameter $r$ takes another meaning, and in
particular can have a coupling-dependent scale.
The long distance expansion
of the monopole problem then becomes a weak coupling expansion,
and the exponentially small corrections can be truly identified
with instanton effects. It will be our goal to interpret these effects
in the various cases where the Atiyah-Hitchin manifold provides the
exact answer, in the hope of drawing lessons for cases where an
explicit answer is missing. This program has already been carried out
at the one-instanton level in the context of three-dimensional
gauge theories \cite{Dorey1,Dorey2}: our goal is to extend this
study to all higher order non-perturbative contributions, and
to other settings where the same effects appear.

The plan of this paper is as follows. We will first review the
various instances of the Atiyah-Hitchin manifold in supersymmetric
gauge theories, brane constructions and string backgrounds, find
the relevant instanton configurations and identify the parameters
$r$ and $\sigma$ in these settings. In Section 3, we will revisit
the Atiyah-Hitchin metric in a way that allows us to easily
extract the series of exponential corrections to the Taub-NUT
limit, and identify precisely which instanton configurations
contribute to which components of the metric. In Section 4, we
shall justify our claim that the exponential corrections arise as
contributions from instanton--anti-instanton bound states, and
discuss the consequences of this phenomenon for the general
question of non-perturbative corrections to hyperk\"ahler
manifolds. Some computational details pertaining to Section 3 are
relegated to the appendices.

\section{Atiyah-Hitchin Manifold, a Festival}
In this section, we would like to review some of the many
instances of the Atiyah-Hitchin manifold in field or string-theoretic
situations, where it provides an exact resummation of all quantum
corrections. Being the moduli space of two $SU(2)$ monopoles,
it naturally arises whenever we embed monopole solutions in string theory.
For a recent discussion on this point see \cite{HZmonopoles}.
By duality, it also appears in many other situations where the
relevance of monopoles is not immediately obvious. Finally,
being a hyperk\"ahler manifold, it appears in many
backgrounds with high degree of supersymmetry.
Our aim here will be to understand the source of non-perturbative
effects, and in particular identify the weight of the semi-classical
configurations in terms of the monopole variables (up to factors
of 2 and $\pi$, which take care of themselves). In the course
of our discussion, we will also mention the occurrence of higher
monopole moduli spaces. Even though they are not as well understood
as the two $SU(2)$ monopole case, there is yet a considerable amount
of knowledge about them which can be carried over to these dual
situations.

\subsection{Brane lifting of the monopole problem}
\label{brane} Much insight into the dynamics of gauge theories has
been gained by embedding them into string theory. The monopole
problem is no exception, and can be given a simple brane
realization. In the limit of far separation, a $k$-monopole
configuration of an $SU(2)$ gauge theory is represented by $k$
oriented D-strings stretched between two D3-branes of Type IIB
theory. That a D-string ending on a D3-brane acts as magnetic
source follows from the worldvolume anomalous coupling $\int
F\wedge B_{RR}$ on the D3-brane worldvolume
\cite{Strominger:1996ac}. The four scalars associated
to each monopole correspond to the three spatial coordinates of
the D1-brane on the D3-brane,
together with a fourth scalar $\sigma$ measuring the zero-mode of the U(1)
gauge field $\int A$ on the stretched D1-brane world-line.

\EPSFIGURE{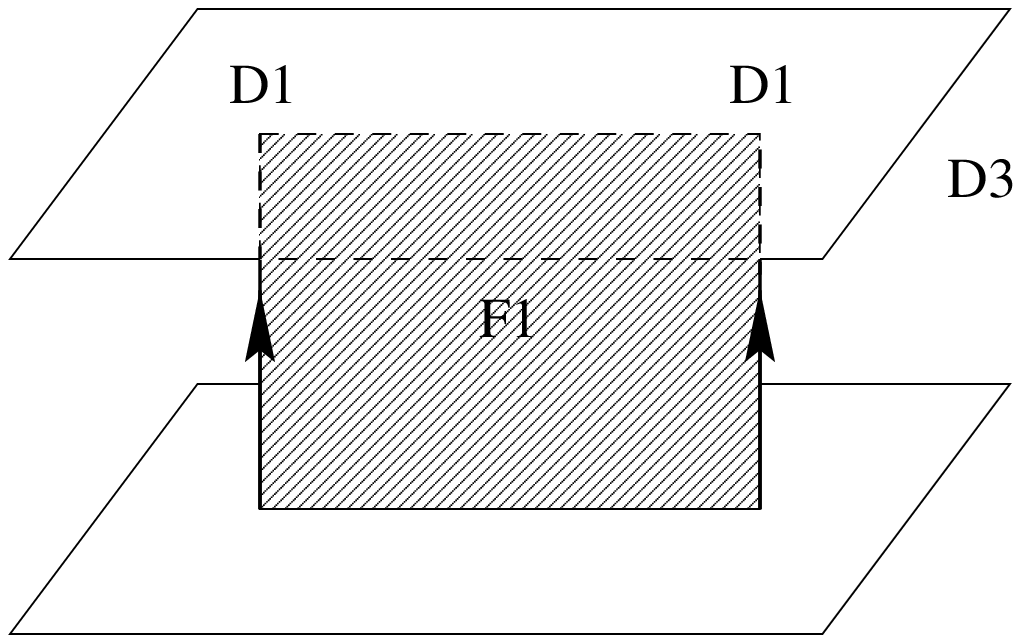,height=4.2cm}{Worldsheet instanton between
two D1-monopoles on a D3-brane worldvolume \label{d3d1f1}}

Monopoles of $SU(N)$ can be similarly represented as D-strings
stretching between different foils of a stack of $N$ D3-branes.
Monopoles in 3+1 dimensions can also be lifted to higher extended objects
in higher dimensions, or reduced to instantons in 3 dimensions,
and so the configuration generalizes to $k$ D$p$-branes stretched
between $N$ D$(p+2)$-branes, for $p=0\dots 6$. One virtue of
this description is that it makes obvious Nahm's construction of
monopoles, by switching the perspective from the D3-brane to the
D1-brane worldvolume: the matrices appearing in Nahm's equations
simply describe the fluctuations of the transverse positions of
the D1-branes stretched between the D3-branes \cite{Diaconescu}.

Another virtue of this representation is that it gives a simple
geometric representation
of the non-perturbative contributions appearing
in \eqref{1inst}: the value of
the Higgs field $\phi$ being related to the distance $L$
between the two D3-branes through $\phi=L/l_s^2$,
the weight $e^{-\phi r}$ simply corresponds to the action
of an Euclidean fundamental string stretched between the two D1-branes and
D3-branes as depicted on Figure \ref{d3d1f1}. We shall refer to this
configuration as a worldsheet instanton. Since the fundamental string
stretched between two D3 is the massive $W$-boson of the gauge theory,
this rephrases our previous statement that the exponential corrections
in \eqref{1inst} come from Euclidean worldlines of $W$-bosons stretching
between the two monopoles. The imaginary part $e^{i\sigma}=
e^{i(\sigma_1-\sigma_2)}$ of
the instanton weight comes from the electric coupling of the boundaries of
the fundamental string worldsheet to the U(1) gauge fields 
$\sigma_1$ and $\sigma_2$ living
on the two D1-branes.



\subsection{Monopoles and three-dimensional gauge theories\label{gauge}}

\EPSFIGURE{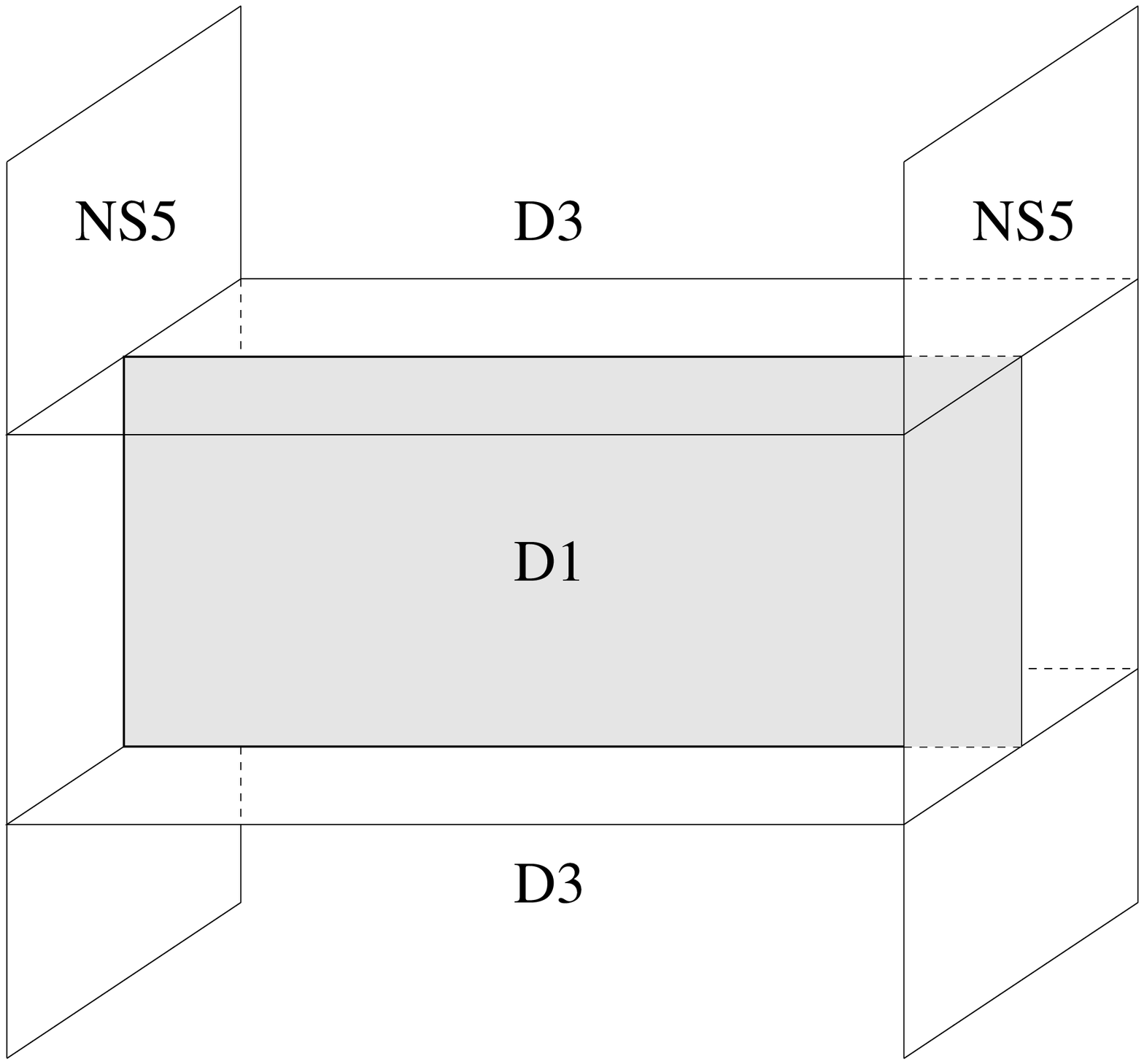,height=6cm}{Duality between monopole
moduli space and three-dimensional gauge theory moduli space\label{d3ns5}}

The above brane configuration for $p=3$ can be
S-dualized to a configuration of $k$ D3-branes stretched between
$N$ NS5-branes. This system was studied in detail in \cite{HW} as
a way of deriving the conjectured equivalence between moduli
spaces of monopoles and the Coulomb branch of three-dimensional
gauge theories with 8 supersymmetries \cite{CH}. Indeed,
the theory living on the D3-branes is effectively at distances
larger than the 5-brane separation a three-dimensional gauge
theory with three Higgs fields in the adjoint (the extra four
Higgs fields living on the D3-branes in vacuo are projected out by
the boundary conditions imposed by the NS5-branes). In the Coulomb
phase, the three-dimensional gauge fields can be dualized into
pseudoscalars $\sigma_i$, which combine with the three Higgs
fields to make $k$ hypermultiplets taking value in some
hyperk\"ahler manifold. For $(N,k)=(2,2)$, it was shown on
symmetry grounds that the only possibility is the Atiyah-Hitchin
manifold \cite{SW}. The fact that the moduli space of two $SU(2)$
monopoles appears as the moduli space of the three-dimensional
gauge theory is naturally explained from the brane point of view,
by simply switching the perspective from the NS5 to the D3-brane
worldvolume \cite{HW}. More generally, the quantum corrected
moduli space of the three-dimensional $SU(k)$ gauge theory is
identified with the moduli space of $k$ monopoles in $SU(2)$
\cite{CH}, and a similar equivalence also holds between moduli
spaces of monopoles in ADE gauge groups and Coulomb branches of
quiver three-dimensional gauge theories \cite{Tong:1999fa}.

{}From the point of view of the three-dimensional gauge theory, the $1/(r-2)^3$
contribution to the Riemann tensor $R_{(0)}$ appears as a one-loop
effect, while the non-perturbative corrections arise as instanton
effects from monopoles. This can easily be seen by following the
instanton configuration on Figure \ref{d3d1f1} under S-duality: it
becomes an Euclidean D1-brane stretched between both the pair of
D3 branes and the pair of NS5-branes, as first discussed in
\cite{HW} (see Figure \ref{d3ns5}). From the point of view of the
D3-brane, this is an Euclidean monopole, and hence an instanton in
the three-dimensional gauge theory. Its classical action is given
by $r L /(g_s l_s^2)$, where $L=\phi l_s^2$ is the distance
between the D3-branes, and $r$ the distance between the
NS5-branes. This can be rewritten in terms of the gauge theory
variables as $\phi / g^2_{YM}$, which is the appropriate weight
for an instanton of the three-dimensional gauge theory. As shown
by Polyakov \cite{Polyakov}, instanton contributions should
in addition be weighted by a term $e^{i\sigma}$ where $\sigma$ is the
dual of the gauge field in three dimensions, or equivalently the
fourth component of the dual magnetic gauge field $A_{m}$ in four
dimensions. Indeed, this imaginary part of the action naturally
arises from a magnetic coupling $\int A_{m}$ on the D1-brane
boundary, dual to the more familiar electric coupling $\int A$ on
the fundamental string boundary.

If the moduli space of the $N=4$ three-dimensional gauge theory is
really the Atiyah-Hitchin manifold, it should be possible to
recover the exponential correction in \eqref{1inst} from a
one-instanton computation. This was carried out successfully in
\cite{Dorey1} for the case without matter, and in \cite{Dorey2}
for the case with one hypermultiplet, which is conjectured to be
described by a double cover of the Atiyah-Hitchin manifold
\cite{SW}. We will be interested in the extension of these
considerations to higher order in the instanton expansion.

\subsection{Orientifold Eight-planes with NS5-branes}

\EPSFIGURE{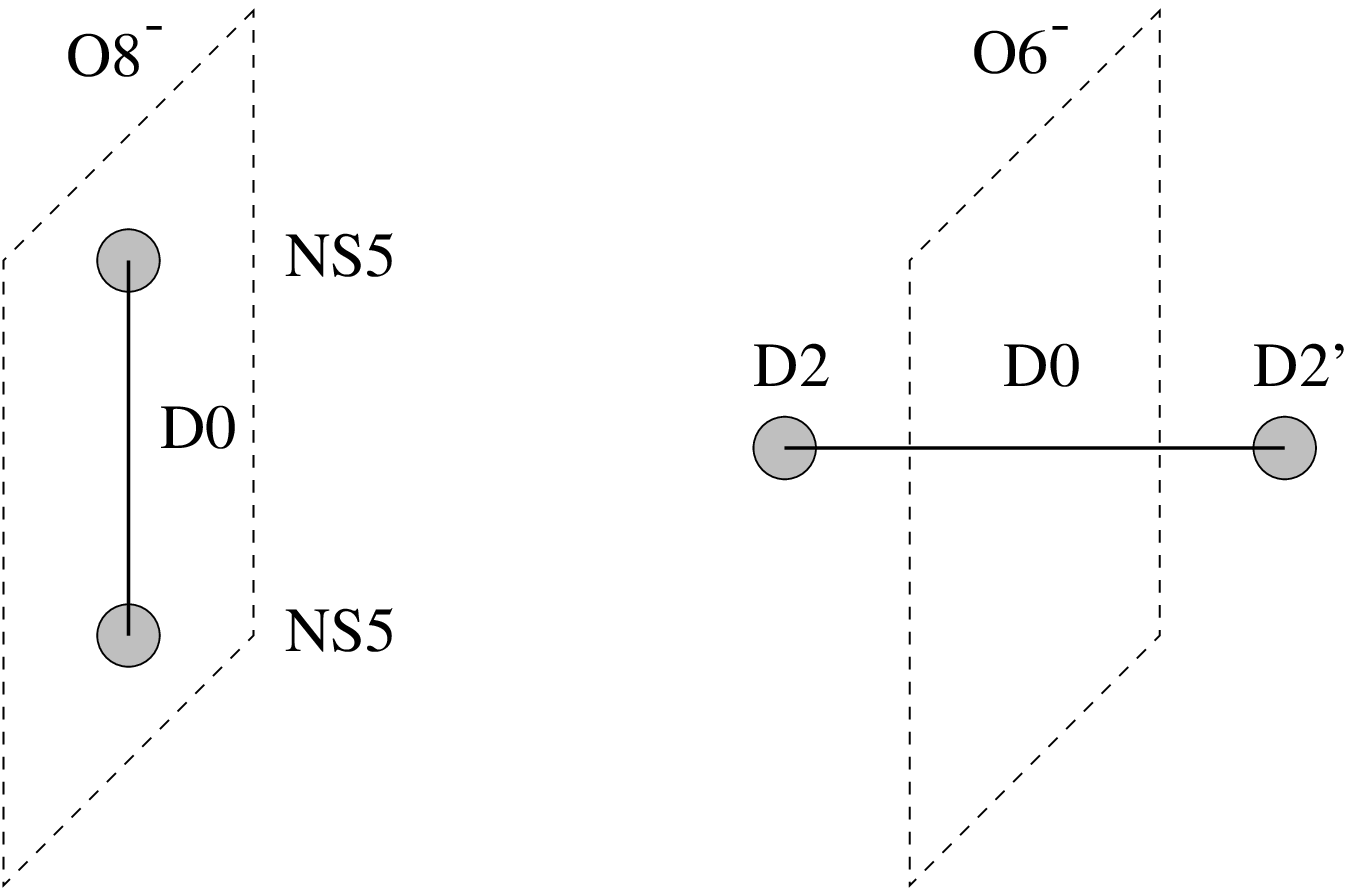,height=6cm}{{\it Left:} $SU(2)$ monopoles seen
as NS5-branes stuck on an orientifold 8-plane at strong coupling.
The instanton corrections come from Euclidean D0-branes stuck on
the orientifold. {\it Right:} Orientifold 6-plane probed by a
D2-brane and its image. The D2-branes are the monopoles of a
Higgsed $SU(2)$ and interact by exchange of D0-branes.
\label{o8o6}.}

Let us now go back to the case of the D$p$-D$(p+2)$ system, for
$p=6$, now embedded in Type I$'$. Enhanced $SU(2)$ gauge symmetry
occurs when two D8-branes become coincident, but also when the
string coupling on one of the $O8^-$ orientifold planes becomes
infinite. The $SU(2)$ symmetry is Higgsed at finite string
coupling, and the massive $W$-bosons are provided by D0-branes stuck
on the orientifold plane. It should also be possible to describe
$SU(2)$ monopoles on an $O8^-$ orientifold at finite coupling.
This problem was discussed in \cite{HZmonopoles}, where it was
shown that the monopoles are NS5-branes stuck on the orientifold
plane (see Figure \ref{o8o6} left). These have a relative three-
dimensional distance with a phase angle corresponding to the
distance between the NS branes in the eleventh direction (string
coupling direction). These four scalars form a hypermultiplet
which is again parameterizing the AH manifold \cite{HZmonopoles}.
The monopoles interact by exchange of $W$-bosons, hence by D0-brane
instantons stretched between the two NS5-branes. The
non-perturbative corrections to the moduli space are thus
suppressed by $\exp[-r/(g_s l_s)]$, where $r$ is the distance
between the NS5-branes. This effect is non-perturbative in the
string coupling.

\subsection{Heterotic string on an ALE space}

One can now T-dualize this configuration to a Type I background
and further S-dualize to a Heterotic string background. The
NS5-branes turn into an $\Real^4/\Zint_2$ singularity and the
corresponding background is the Heterotic string on $\Real
^{5,1}\times ALE$, where $ALE$ is the Eguchi-Hanson manifold which
resolves the orbifold singularity. The hypermultiplet controlling
the size and B-flux of the blown-up two-cycle can be argued to
take value again in Atiyah-Hitchin manifold \cite{Sen}. The
origin of the exponential corrections can be traced by following
the sequence of dualities, and correspond to world-sheet
instantons of the fundamental Heterotic string. This is in
agreement with the non-renormalization property of the
hypermultiplet moduli space in Heterotic theories with 8
supersymmetries. The exponential corrections are of order
$\exp(-A/l_s^2+i B)$, where $A$ denotes the area of the two-sphere
in the blown-up ALE space and $B$ is the B-flux on that cycle.
Since these corrections occur purely at string tree-level, it
should be possible to recover the exact Atiyah-Hitchin metric in a
sigma-model computation. This is however not easy due to the
singular nature of the conformal field theory at hand. Note that
the identification of Heterotic hypermultiplet spaces and monopole
moduli spaces has been generalized in
\cite{Witten:1999fq,Rozali:1999va}.

\subsection{D6-branes on $K3$}
In a recent series of papers \cite{JPP, Jarv, Johnson} a yet
different type of realization of the Atiyah-Hitchin manifold was
considered. One looks at a collection of $N$ D6 branes wrapped on
a smooth $K3$. The low energy dynamics of these branes is given by
a 2+1 dimensional gauge theory with 8 supercharges and gauge group
$U(N)$. For the special case of $N=2$ this configuration leads to
the gauge theory which was discussed in subsection \ref{gauge} and
hence the moduli of the wrapped D6 branes parameterize the
Atiyah-Hitchin manifold. The gauge coupling of the 2+1 dimensional
theory can be computed to be ${1}/{g_3^2}=(V_{K3}-l_s^4)/(g_s
l_s^3)$, where the negative term comes from extrinsic curvature
terms on the D6-brane. It vanishes at $V_{K3}=l_s^4$, which is a
point of $SU(2)$ enhanced symmetry in target space (as seen from
duality with the Heterotic side for example). The D6-branes can be
seen as the monopoles of this broken gauge symmetry in 5+1
dimensions, as they become tensionless at the enhanced symmetry
point. The Higgs vev of the three-dimensional gauge theory on the
D6-brane is related to the distance $r$ between the 2 D6-branes by
$\phi={r}/{l_s^2}$. The expansion parameter controlling the
corrections to the D6-brane moduli space is then $r
(V_{K3}-l_s^4)/(g_s l_s^5)$. This is also the action of Euclidean
D4-branes wrapped on $K3$ and stretching between the D6-branes,
which are therefore the relevant instanton configurations. This is
consistent with the identification of the wrapped D4-branes as the
$W$-bosons of the target-space $SU(2)$ enhanced symmetry. This also
identifies the Higgs vev of the target-space gauge symmetry as
$(V_{K3}-l_s^4)/(g_s l_s^5)$. In fact, the statement made in
\cite{JPP, Jarv, Johnson} is slightly different, since it is
concerned with the moduli space of a probe D6-brane in the
background of a large number $N$ of D6-branes creating a repulson
singularity. The claim is that the corrected moduli space is again
the Atiyah-Hitchin manifold. The latter then appears as a
four-dimensional submanifold of the moduli space of a large number
$N$ of $SU(2)$ monopoles.


\subsection{Atiyah-Hitchin Backgrounds}

An interesting occurrence which is different from all examples
discussed above is a string/M-theory background on a manifold
which contains the Atiyah-Hitchin manifold, $M_{AH}$. A simple
example for such a background is given by M-theory on
$\Real^{6,1}\times M_{AH}$. This background is known to be the strong
coupling dual of Type IIA string theory on $\Real^{6,1}\times \Real^3/I_3
\times\Omega\times(-1)^{F_L}$ or in its more familiar form, an
$O6^-$ plane \cite{SW}. One may argue that D2-brane probes in the
vicinity of an orientifold 6-plane behave as monopoles of an
Higgsed $SU(2)$ gauge symmetry (see Figure \ref{o8o6}, right).

Indeed, the worldvolume theory on a pair of D2-branes is a pure
three-dimensional
$N=4$ $SU(2)$ theory, and hence has the Atiyah-Hitchin manifold as
its moduli space. It should therefore be the case that the
singular O6-plane be resolved in the strong string coupling limit
into a smooth manifold, the Atiyah-Hitchin manifold itself. In
order to relate the Higgs vev to the string parameters, let us
consider the gauge theory on the D2-branes. The three dimensional
gauge coupling is given by the usual ${1}/{g_3^2}={l_s}/{g_s}$.
Denote by $r/2$ the distance of the D2 brane to the $O6^-$ plane,
measured by the $SU(2)$ vev $\phi_{D2}={r}/{l_s^2}$. The
exponential corrections are given by Euclidean D0-branes which are
stretched in between the D2 brane and its image with an expansion
parameter ${\phi_{D2}}/{g_3^2}={r}/{g_s l_s}$. This implies that
the $SU(2)$ gauge theory of which the D2-branes are monopoles will
have a Higgs vev $\phi_{O6}={1}/{g_sl_s}$. The enhanced $SU(2)$
symmetry therefore occurs at scale $\phi_{O6}$, which is also the
radius of the eleventh dimension. The $W$-bosons of the
enhanced $SU(2)$ symmetry are the D0-branes, {\it i.e.} the momentum
modes of the graviton on the compact eleventh dimension.

\section{Atiyah-Hitchin revisited}
Having recalled a few occurrences of the Atiyah-Hitchin manifold
in  string and field theory, and identified what type of
non-perturbative corrections it purports to resum, we now would
like to extract the precise form of these corrections beyond the
one-instanton effect that was displayed in \eqref{1inst}. For
convenience, we shall use the monopole terminology, but the other
cases can be obtained by simply reinterpreting the meaning of the
$r$ and $\sigma$ coordinates.

\subsection{The Atiyah-Hitchin metric and modular forms}
The metric found by Atiyah and Hitchin was originally expressed in
terms of elliptic functions, whose appearance is hardly surprising
given the fact that the algebraic curve underlying the Nahm
equations has genus 1 in the two-monopole case. The asymptotic
expansion of elliptic functions is most easily obtained after
expressing them in terms of Jacobi Theta functions and other
Eisenstein series, whose $q$-expansion is well known. It turns out
that the metric can be expressed very concisely in that form, as
we now briefly show \footnote{After obtaining this result, we were
informed by I. Bakas that it had already appeared in the
mathematical literature \cite{takh}. }. We follow the notations
and conventions of \cite{Gibbons:1986df} for the Atiyah-Hitchin
metric, and of \cite{Kiritsis} for modular forms.

The general $SO(3)$ invariant four-dimensional metric can
be chosen in the Bianchi IX form,
\begin{equation}
\label{ds}
ds^2=(abc)^2 dt^2+a^2\sigma_1^2 + b^2\sigma_2^2 + c^2\sigma_3^2
\end{equation}
where $\sigma_{1,2,3}$ are a basis of $SO(3)$-invariant one-forms
fulfilling the algebra
\begin{equation}
d\sigma_i=\frac{1}{2}\epsilon_{ijk}\sigma_j\wedge \sigma_k.
\end{equation}
An explicit representation for these one-forms is given in
terms of the Euler parameterisation of $SU(2)$ as
\begin{eqnarray}
\sigma_1 &=& \sin\psi d\theta -\cos\psi\sin\theta d\phi, \nn\\
\sigma_2 &=& -\cos\psi d\theta -\sin\psi\sin\theta d\phi, \\
\sigma_3 &=&  d\psi+\cos\theta d\phi\nn
\end{eqnarray}
The ranges of $\theta$, $\phi$ and $\psi$ are $[0,\pi]$, $[0,2\pi]$ and
$[0,2\pi]$ respectively, up to identifications to be discussed below.
Requiring the curvature to be self-dual puts three constraints on the
coefficients $a,b,c$,
\begin{equation}
\label{sdu}
\frac{a'}{abc}=  \frac{b^2+c^2-a^2}{2bc} - \lambda\ ,\ \mbox{etc }
\end{equation}
plus the two others obtained by cyclic permutation of $a,b,c$.
Here, the prime denotes differentiation with respect to the radial
parameter $t$ (see below equation \ref{rt} for a relation between
$r$ and $t$.) and $\lambda$ is an integration constant which is
set to 1 in the Atiyah-Hitchin case. Following
\cite{Atiyah:1985dv}, we define $w_1=bc,w_2=ca,w_3=ab$ to rewrite
the differential system as
\begin{equation}
\label{dw}
d(w_1+w_2)/dt=-2 w_1 w_2 \ ,\ \mbox{etc}
\end{equation}
known as the Halphen system. Now we observe that Jacobi Theta functions
give a simple solution of that system, since they fulfill the modular
identity
\begin{equation}
\frac{d}{dt}\left( \frac{\th_3'}{\th_3} + \frac{\th_4'}{\th_4} \right)
= \frac{2}{\pi} ~ \frac{\th_3'}{\th_3} \cdot \frac{\th_4'}{\th_4}
\end{equation}
where the complex modulus of the Theta function is $\tau=i t$ \footnote{The
relation of modular forms to Halphen-like differential systems
has been discussed in \cite{harnad}.}.
This implies that a solution of \eqref{dw} can be chosen as
\begin{eqnarray}
\label{ww123}
w_1&=&\frac2\pi \frac{\th_2'}{\th_2}=
-\frac{\pi}{6}\left(E_2+\th_3^4+\th_4^4\right)\ \nn\\
w_2&=&\frac2\pi \frac{\th_3'}{\th_3}=
-\frac{\pi}{6}\left(E_2+\th_2^4-\th_4^4\right)\ ,\\
w_3&=&\frac2\pi \frac{\th_4'}{\th_4}=
-\frac{\pi}{6}\left(E_2-\th_2^4-\th_3^4\right)\ ,\nn
\end{eqnarray}
which will be shown to satisfy the appropriate boundary
conditions. The second equality on the righthand side follows from
a standard equality involving the holomorphic Eisenstein series
$E_2$. Note that $w_1<0,w_2<0$ and $w_3>0$, or equivalently
$a>0,b>0,c<0$. The relation to Atiyah and Hitchin's original
formulae is detailed in Appendix \ref{AAA}.

The above elliptic functions involve two different asymptotic
regimes, $\tau\to i\infty$ and $\tau\to 0$, corresponding to the
coincident limit and the large separation limit of the monopoles,
respectively. Indeed, as $t\to \infty$, $w_2$ and $w_3$ become
equal and the metric takes the form
\begin{equation}
ds^2=4\pi^2 e^{-2\pi t} \left( \pi^2  dt^2 + 4 e^{-2\pi t} \sigma_1^2\right)
+\frac{\pi}{2} \left( \sigma_2^2+\sigma_3^2 \right)
\end{equation}
up to exponentially small corrections.
Changing variables to $\tilde u=e^{-\pi t}$, this is recognized as
a bolt singularity, which is a mere coordinate singularity if we
take the quotient by the symmetry $I_1:
\theta\to\pi-\theta,\phi\to\phi+\pi,\psi\to-\psi$. The
resulting space is a double cover of the Atiyah-Hitchin manifold,
the latter being obtained after modding out in addition by
$I_3:\psi\to \psi+\pi$.

In the $t\to 0$ limit, it is necessary to perform a modular transformation
$w_i(t)=-\frac{1}{t}-\frac{1}{t^2} w_j(1/t)$,
where $j=3$ for $i=1$; $j=1$ for $i=3$; and $j=2$ for $i=2$. This yields
\begin{eqnarray}
\label{rt}
w_1(t)&=&-\frac{1}{t} - \frac{\pi}{t^2}(4 q + 8 q^2
+\dots )\sim -r/\pi\nn\\ w_2(t)&=&-\frac{1}{t} + \frac{\pi}{t^2}(4
q - 8 q^2 +\dots )\sim -r/\pi\\ w_3(t)&=&-\frac{1}{t} +
\frac{\pi}{t^2}(\frac12+4 q^2 + \dots ) \sim r^2(1-2/r)/(2\pi)\nn
\end{eqnarray}
where $q=e^{-\pi/t}$ and we defined $r=\pi/t$.
The first term in $1/t$ arises because of the
anomalous modular property of the Eisenstein series $E_2$.
The Atiyah-Hitchin metric \eqref{ds} thus reduces to
\begin{equation}
ds^2=\frac{1}{2\pi}\left[ \left(1-\frac{2}{r}\right)
\left( dr^2+ r^2 (\sigma_1^2+\sigma_2^2) \right)
+ \frac{4}{1-\frac{2}{r}} \sigma_3^2 \right]
\end{equation}
where we recognize the Taub-NUT metric with mass parameter $-1$.
In particular, the asymptotic geometry is $\Real^3\times S_1$
(mod $\Zint_2$) and $r$ can be identified as the distance between
the monopoles, in units of the $W$-boson mass. 
The angle $\psi$ parameterizes the circle $S_1$
and is the coordinate that we called $\sigma$ before; giving
momentum in that direction amounts to giving electric charge to
the monopole, turning it into a dyon.

\subsection{Four-fermion terms and curvature}
We now would like to extract the exponential corrections from the
metric described above. For this purpose, we find it convenient to
use the language of the three-dimensional $SU(2)$ gauge theory
with 8 supercharges. As described in Subsection \ref{brane}, the
three Higgs scalars combine with the pseudoscalar $\sigma$ dual to
the gauge field in three dimensions to make an hypermultiplet
taking values in the Atiyah-Hitchin manifold. The $N=4$ gauge
theory has a symmetry group $SU(2)_H\times SU(2)_V \times SU(2)_E$
which is the product of the R-symmetry $SU(2)_H$ already present
in the six-dimensional $N=1$ theory, the R-symmetry $SU(2)_V$
coming from compactification from 6 to 3 dimensions, and the
Euclidean group in three dimensions. The fermions transform as
$(2,2,2)$ and the bosons as $(1,3+1,1)$, so that $SU(2)_V$ is
broken to a $U(1)_V$ subgroup by the Higgs vev. We choose to align
the Higgs field along the ``vertical'' direction, $\theta=0$.

Given that the theory has 8 supersymmetries, the first quantum corrections
arise in the metric of the scalars, or the four-fermion interactions
which are related to the former by supersymmetry. It is thus convenient
to concentrate on the four-fermion terms, which are contracted with the
Riemann tensor of the bosonic moduli space. The antisymmetric product
of four fermions transforms as
\begin{equation}
(5,1,1)+(1,5,1)+(1,1,5)+(3,3,3)+(3,3,1)+(3,1,3)+(1,3,3)+(1,1,1) \nn
\end{equation}
out of which we must keep the singlets under the Euclidean group
$SU(2)_E$ and the R-symmetry group $SU(2)_H$, since the scalars
are neutral under them. This only leaves the $(1,5,1)+(1,1,1)$
component.

On the other hand, the Riemann tensor of a hyperk\"ahler manifold
transforms as $(1,5,1)$. Indeed, since the Riemann tensor is
self-dual, the independent components are $R_{i;j}:=R_{0i0j}$
which make a symmetric tensor of $SU(2)$, and its trace is zero by
the cyclic property of the Riemann tensor. It is therefore
contracted in the effective action with the $(1,5,1)$ part of the
4 fermion product only. Since the Higgs vev breaks $SU(2)_V$ to
$U(1)_V$, we can split the fluctuations of the three scalars into
a complex field $w=x+iy$ of $U(1)_V$ charge $++=1$, its complex
conjugate $\bar w=x-iy$ of $U(1)_V$ charge $--=-1$, and a real scalar
$z$ of charge $+-=0$. Taking into account the change of basis, the
Riemann tensor decomposes into
\begin{eqnarray}
R_{(2)}&=&R_{++;++}=e^{i\sigma} (R_{1;1}-R_{2;2}+ 2 i R_{1;2})\nn\\
R_{(1)}&=&R_{++;+-}=e^{i\sigma/2}(R_{1;3}+i R_{2;3})\nn\\
R_{(0)}&=&R_{++;--}-R_{+-;+-}=R_{1;1}+R_{2;2}-R_{3;3}\\
R_{(-1)}&=&R_{--;+-}=e^{-i\sigma/2}(R_{1;3}-i R_{2;3})\nn\\
R_{(-2)}&=&R_{--;--}=e^{-i\sigma}(R_{1;1}-R_{2;2}- 2 i R_{1;2})\nn
\end{eqnarray}
These components hence appear in the effective action contracted
with the fermion quadrilinears in such a way that the $SU(2)_H\times
U(1)_V\times SU(2)_E$ quantum numbers are trivial. In particular,
since $\sigma$ enters in the theory only through exponential
effects $e^{i n\sigma}$, with $n$ integer, it should be the case
that $R_{(1)}=R_{(-1)}=0$ exactly, and $R_{(2)}=R_{(-2)}=0$ in
perturbation theory. We shall shortly see that this indeed holds.

\subsection{The non-perturbative expansion of the Riemann tensor}
Having identified the precise components of the Riemann tensor to
which instanton effects contribute, we can now proceed to evaluate
them using the modular expression of the Atiyah-Hitchin metric. We
use the orthonormal basis $e_0=abc~dt,e_i=a_i\sigma_i$. In terms
of the parameters $a,b,c$ entering \eqref{ds}, we find (see
Appendix \ref{BBB} and \cite{Gauntlett:1996fu})
\begin{equation}
R_{1010}=\frac{a'}{a(abc)^2}\ ,\quad R_{2020}=\frac{b'}{b(abc)^2}\
,\quad R_{3030}=\frac{c'}{c(abc)^2}\ ,\quad R_{i0j0}=0
\quad\mbox{if}\quad i\neq j\ .
\end{equation}
Using the differential system \eqref{sdu}, this can be
re-expressed in terms of the $w$'s as
\begin{equation}
R_{1010}=\frac{1}{w_2 w_3}\frac{d}{dt} \left(
\frac{(w_1 w_3)^2+(w_1 w_2)^2-(w_2 w_3)^2}{w_1^2 w_2 w_3} \right)\
\mbox{etc.}
\end{equation}
and expressed in terms of modular forms using \eqref{ww123},
\begin{eqnarray}
R_{1010}&=&\frac{1}{(E_2+e_3)(E_2+e_4)}\times \\
&&\frac{d}{dt}\left[ \frac{E_2^4 + 4 E_2^3 e_2 - 2 E_2^2(e_2^2 + 2
e_3 e_4) - 4 E_2 E_6 + e_2^4 - (e_3 e_4)^2 - 4 e_2 E_6}
{(E_2+e_2)^2(E_2+e_3)(E_2+e_4)}\right]. \nonumber
\end{eqnarray}
The modular forms $e_2,e_3,e_4$ are the roots of the polynomial
$x^3-3E_4 x -2E_6$ and are defined in Appendix \ref{AAA}. This
result, although not very enlightening, has the virtue of
showing that $R_{1;1}$ is a modular form invariant under a
$\Gamma^-(2)$ subgroup of $Sl(2,\Zint)$ exchanging $e_3$ and $e_4$
but leaving $e_2$ invariant. Similarly $R_{2;2}$ is invariant
under $\Gamma^0(2)$ and $R_{3;3}$ under $\Gamma^+(2)$, while
general modular transformations permute these groups. An important
consequence is that {\it $R_{1;1}$ and $R_{2;2}$ have the same
$q$-expansion up to alternating signs}. Hence {\it the expansion
of $R_{(0)}$ only involves even powers, while the expansion of
$R_{(\pm 2)}$ only involves odd powers}. As we shall see shortly,
this is an important consistency check on the interpretation of
the result as coming from instanton--anti-instantons bound states.

More precisely, keeping into account the anomalous modular 
transformation of $w_i(t)$, we find that the Riemann tensor 
has a large distance expansion 
\begin{eqnarray}
\label{Rq}
R_{(0)}&=&\frac{4}{(r-2)^3}-  \sum_{n=1}^{\infty} 
\frac{P_{2n}(r)}{(r-2)^{n+3}} q^{2n} \\
R_{(\pm2)}&=& \sum_{n=0}^{\infty} 
\frac{P_{2n+1}(r)}{(r-2)^{n+2}} q^{2n+1} e^{\pm i\sigma}
\end{eqnarray}
where $q=e^{-r}$, and 
$P_n(r)$ is a polynomial of order $3([n/2]+1)$ in $r$ 
with alternating integer coefficients. 
The first terms are easily computed using 
Mathematica,
\begin{eqnarray*}
P_1(r)&=&8(-3+9r-6r^2+r^3)\\
P_2(r)&=&32(6-54 r+123r^2-124r^3+66r^4-18r^5+2r^6)\\
P_3(r)&=&32(6-117r+459r^2-702r^3+495r^4-162r^5+20r^6)\\
P_4(r)&=&64(-6+ 333\ r - 2811r^2 + 9023r^3 - 14928r^4 + 14352r^5 -
    8392r^6 + 2952r^7 \\ && - 576r^8 + 48r^9)
\end{eqnarray*}
The leading $r$ power can be extracted at each
order in $q$, yielding
\begin{eqnarray}
\label{Rqr}
R_{(0)}
&=&\frac{2^2}{r^3}-2^6 r^2 q^2-3 \cdot 2^{10}  r^4 q^4 -3 \cdot2^{15}  r^6 q^6
-5\cdot  2^{19} r^8 q^8- 3\cdot 7 \cdot2^{22} r^{10} q^{10}+\dots \nn\\
R_{(\pm 2)}
&=&e^{\pm i\sigma}\left(2^3 r q + 5 \cdot 2^7 r^3 q^{3}+5 \cdot
2^{13} r^5 q^{5}
+11 \cdot 2^{16}  r^7 q^{7}+ 5\cdot 7 \cdot 2^{19}   r^9 q^{9}+\dots\right)\nn
\end{eqnarray}
Several remarks are in order about these results.
\begin{itemize}
\item First, as anticipated
in the last section, the perturbative correction only occurs in
$R_{(0)}$, as it should be since $U(1)_V$ is a symmetry preserved
by perturbation theory.
\item Second and most importantly,
exponential corrections come with
an arbitrary power $n$ of the semiclassical weight $q=e^{-r}$, but only
with zero-th power of $e^{i\sigma}$ for $n$ even, or first power
for $n$ odd. This suggests that the correct interpretation of the
non-perturbative exact result is rather a sum of {\it bound states
of $n$ instantons and $\bar n$ anti-instantons}, with overall topological
charge $n-\bar n=0$
for $R_{(0)}$ and $\pm 1$ for $R_{(\pm 2)}$, as appropriate for charge
conservation.
\item Third, each power of $q$ comes in with
an additional power of $(2r)^{3/2} / (r-2)^{1/2}$, which indicates
that the action of the semiclassical configuration is
\begin{equation}
\label{scllog}
S_{cl} = (n+\bar n) \left( r -\frac12 \log\frac{(2r)^3}{r-2} \right)
+ i (n-\bar n) \sigma
\end{equation}
This result agrees at large $r$ with the one-instanton result of
\cite{Dorey1}, who found a contribution proportional to
$\phi e^{-8\pi^2 \phi/g_3^2}$, where the prefactor $\phi$ originates
from the one-loop determinant in the instanton background. Our result
implies higher loop corrections in the one-instanton background. Besides,
it predicts higher order contributions from bound states of instantons and
anti-instantons.
\item We may have considered instead the near-coincident limit $r\to 0$.
The Riemann tensor also admits a $q$-expansion in that regime, with
$q=e^{-1/r}$. In fact, this expansion only involves powers of $q$ and
not of $r$, since no modular transformation is required, and all
coefficients are integer numbers. We do not know of a semi-classical
interpretation of this expansion.
\end{itemize}
This concludes our dissection of the Atiyah-Hitchin metric.
The form of the expansion \eqref{Rq} therefore strongly suggests
the interpretation of the non-perturbative corrections as
contributions of instanton--anti-instanton bound states.
We will now try to give further support for this
unorthodox claim.

\section{Discussion}

Half-BPS saturated couplings in supersymmetric string or field
theories are commonly thought to satisfy some sort of
non-renormalization theorem, restricting them to receive
contributions from a limited order in perturbation theory (usually
one-loop), as well as exponential corrections coming from half-BPS
instantons. This is a well-established fact in a few particular
cases, including the prepotential of $N=2$ gauge theories in 4
dimensions, or some higher-derivative terms in string theory. 
In more general cases, this
expectation is based on a simple zero-mode counting argument: an
$n$-fermion vertex in the low energy effective action can only
receive corrections from instanton configurations with less than
$n$ fermionic zero-modes, and hence breaking at most $n$
supersymmetries. This crude argument would seem to rule out
contributions from bound states of $N$ BPS instantons, which
possess $nN$ zero-modes, but $n(N-1)$ of them are usually lifted
by the fermionic interactions in the action governing the
collective coordinates. Non-BPS instantons (and in particular
superpositions of instantons and anti-instantons) break more
supersymmetries, and hence would seem to have too many fermionic
zero-modes to make any contribution to the half-BPS couplings.

In the case at hand, we are interested in a four-fermion coupling
in a theory with eight supersymmetries. In the language of the
three-dimensional gauge theory, the relevant instanton
configurations are 't Hooft-Polyakov monopoles, which have four
fermionic zero-modes. In the particular case where the three Higgs
fields are aligned (which is automatic in the $SU(2)$ case), the
$N$-monopole configurations are in fact exact solutions (in
contrast to Yang-Mills instantons in four dimensional
gauge theories with a Higgs vev), and hence the $4N$ fermionic
zero-modes cannot be lifted. This argument thus predicts
contributions from only one BPS monopole to the four-fermion
coupling \cite{SW}. Our expansion \eqref{Rq} clearly points to a
flaw in this line of thought. Indeed, we seem to find
contributions from arbitrary numbers of instantons and
anti-instantons at the same time, with a net instanton number
$n-\bar n=0,\pm 1$ \footnote{A crucial check on this
interpretation is provided by the agreement between the net
instanton number and the parity of the semi-classical weight of
the instanton correction.}. The argument above still correctly
predicts the {\it net} instanton number, at least when we focus on
the particular four-fermion vertex $\psi_+\psi_+\psi_+\psi_+$.
This is in fact a simple consequence of $U(1)_V$ charge
conservation, which is violated by $2(n-\bar n)$ units in a
classical background with topological charge $(n-\bar n)$. An
instanton--anti-instanton configuration on the other hand has
vanishing topological and hence $U(1)_V$ charge, and therefore can
be added without disturbing charge conservation.

The argument based on fermionic zero-mode counting can also be
evaded. An instanton--anti-instanton configuration does break all
supersymmetries, but the action of the supercharges on the
instanton configuration does not generate zero-modes of the Dirac
operator, since the configuration is not an exact saddle point of
the action in the first place. It is in the limit of far
separation only, which implies that the fermionic determinant
vanishes in the large distance regime in the space of collective
coordinates. At finite distance, the number of exact zero-modes is
given by the index theorem, which yields $4|n-\bar n|$ independent
of $n$ and $\bar n$ separately \footnote{The instanton field
configuration being not self-dual anymore, the index theorem only
counts the difference between zero-modes of different chiralities,
but one does not expect more zero-modes than the minimum number
$4|n-\bar n|$.}. For $|n-\bar n|=1$, those are saturated by the
four-fermion vertex. We are thus left to evaluate
\begin{equation}
\label{modint}
A=\int_{{\cal M}_{(n,\bar n)}} {\rm Pf}'(D) ({\det}'( \nabla^2))^{-1}
e^{-S_{cl}+i\sigma(n-\bar n)}
\end{equation}
where ${\cal M}_{(n,\bar n)}$ denotes the space of bosonic
collective coordinates of the $(n,\bar n)$
instanton--anti-instanton configuration, ${\rm Pf}'(D)$ the
Pfaffian with the $4|n-\bar n|$ zero-modes deleted, and ${\det}'(
\nabla^2)$ the bosonic fluctuation determinant with zero-modes
deleted. The  space of bosonic collective coordinates is not a
completely well-defined notion, but makes sense in a dilute gas
approximation. The integrand vanishes in the limit of far
separation, but can give a finite value from the bulk of the
moduli space. In fact, our analysis of the Atiyah-Hitchin metric
predicts the value of the integral \eqref{modint} for any $|n-\bar
n|=0,1$. It is interesting to note that the total weight in
\eqref{scllog} is simply $(n+\bar n)$ times the weight of an
individual instanton (which does receive a logarithmic correction
from the quantum fluctuations around it). The effects of the
interactions between instantons and anti-instantons are encoded in
the polynomial factors appearing in \eqref{Rq}, and it would be
interesting to analyze them in more detail. In particular, the
fact that the coefficients are integer, albeit warranted by the
underlying modular symmetry, suggests some topological or counting
interpretation of these coefficients.

\EPSFIGURE{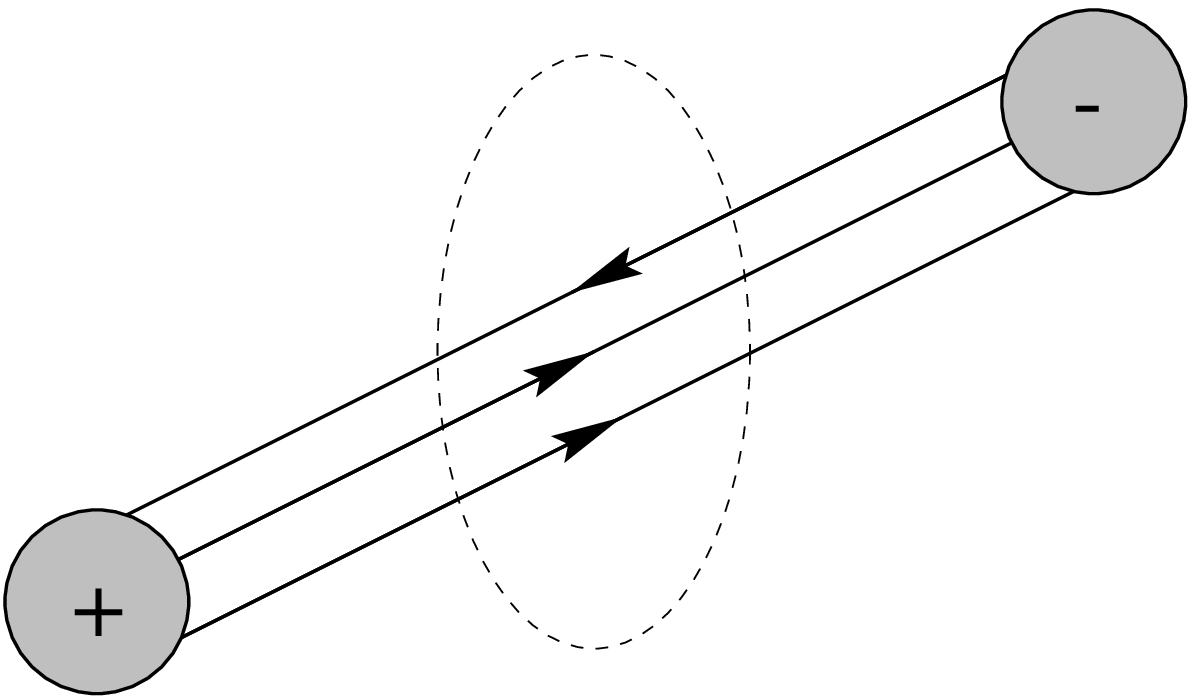,height=4cm}{Bound state of 2 instantons
and one anti-instanton stretching between 2 monopoles\label{2mono}. The
charge across the loop is locally conserved.}

The physical picture that emerges from this discussion is
therefore that the exponential corrections to the monopole moduli
space arise from semi-classical configurations of $2n+1$
fundamental string world-sheets connecting the two D-strings in
the D3-brane setting, or $2n+1$ Euclidean world-lines of $W$-bosons
linking the two monopoles. As depicted in Figure \ref{2mono},
$n+1$ of them are oriented in one way and $n$ in the other, so
that the total charge cancels (taking into account the
net polarization induced by the $\psi_+\psi_+\psi_+\psi_+$ vertex). In
the three-dimension-al gauge theory language, these are $n+1$
instantons and $n$ anti-instantons occurring at arbitrary
Euclidean time.

We can ask how this picture generalizes to more than two
monopoles, where only the large distance behaviour corresponding
to massless exchange is known \cite{Gibbons:1995yw}. Having surmounted the
mental barrier of
supersymmetry, we suggest that the semi-classical configurations
controlling the $n$-SU(2) monopole dynamics are given by strings
connecting monopoles in charge-conserving configurations. This
includes two-particle interactions as in Figure \ref{2mono}, but
also higher point interactions as depicted in Figure \ref{3mono}.
Note that in the three-monopole case, the tantalizing $Y$-shaped
configuration is not a minimum action configuration, since, due to
charge conservation, it is really two $Y$'s with opposite
orientations, which prefer to relax into two counter-rotating
triangles. Of course, these semi-classical contributions are
sub-leading with respect to the power corrections around the
two-monopole interactions due to the presence of a third. These
have been discussed in \cite{Fraser}. It would be interesting to
test our prediction against explicit results for multi-monopoles
moduli spaces, as obtained for instance in \cite{Houghton}.

\EPSFIGURE{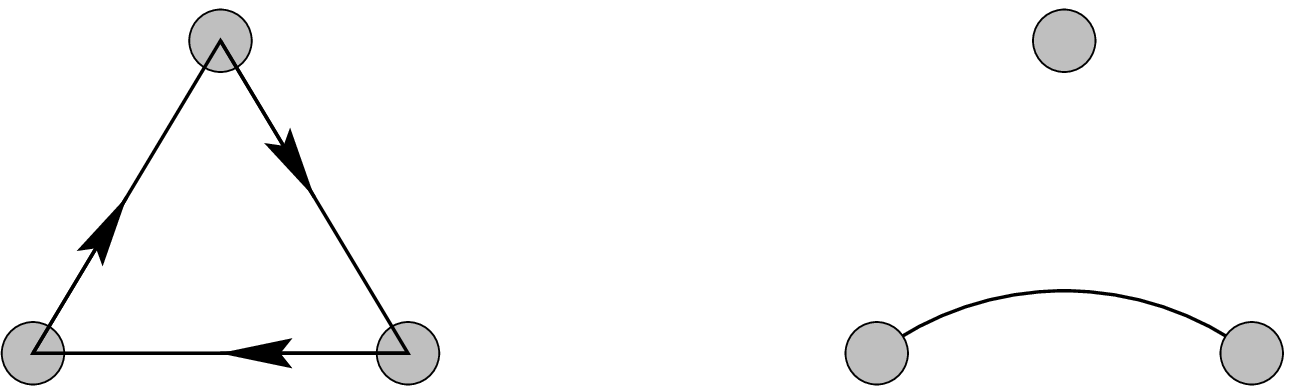,height=3cm}{Sub-leading semi-classical
corrections to the three-monopole moduli space. The two-monopole
interaction is modified by the presence of the third.
\label{3mono} }

It is also natural to ask if it is possible to precisely compute
the higher-instanton effects in \eqref{Rq} from first principles.
Equation \eqref{modint} is first-principled, but seemingly
amenable only to numerical computation. The embedding as a D1-D3
system seems more tractable, but would require some understanding
of stacks of fundamental strings with opposite orientations. The
NS5-brane setting may also offer some interesting insight, since
the instanton is a D-string for which there exists a second
quantized description allowing to consider stacks of them (see for
instance \cite{DVV}). The best bet may actually be the Heterotic
setting, since the instanton corrections simply arise in that case
from world-sheet instantons in the sigma-model $\Real^4/\Zint_2$.
The latter is unfortunately little understood due to the
unresolved singularity. We can however speculate on a possible
solution: since the Atiyah-Hitchin metric arises in the decoupling
limit of the singularity, one may try to approximate the
$\Real^4/\Zint_2$ singularity by a $CP^1$ sigma-model describing
the vanishing two-sphere. Interestingly enough, Cecotti and Vafa
have found long ago similar instanton--anti-instanton
contributions to the topological-anti-topological fusion
coefficients in the $(2,2)$ supersymmetric $CP^1$ sigma model
\cite{Cecotti:1992vb} (the only example so far of such
instanton--anti-instanton effects to our knowledge) \footnote{We
thank C. Vafa for bringing this work to our attention.}. Even more
tantalizingly, the $t\bar t$ equation controlling the $CP^1$
fusion coefficients is nothing but the $SU(2)$ Toda equation
\cite{Cecotti:1992vb}, while the Atiyah-Hitchin is also known to
be controlled by an $SU(\infty)$ Toda equation (see
\cite{Bakas:2000wq} for a review). The study of $(4,0)$
supersymmetric sigma-models may therefore shed an interesting new
light on the corrections to hypermultiplet manifolds.

\newpage

\acknowledgments
We are grateful to S. Cherkis, M. Douglas, R. Gopakumar, M. Gutperle, K. Hori,
D. Tong, C. Vafa for valuable discussions, and especially to I.
Bakas for correspondence and useful guidance into the literature.
We would also like to thank the organizers of the workshop TMR
2000 in Tel Aviv, 7-11 January 2000, where this work was initiated,
for their kind invitation and financial support. B.P. is supported
in part by DOE grant DE-FG02-91ER-40654. A. H. is partially
supported by the DOE under grant no. DE-FC02-94ER40818, by an A.
P. Sloan Foundation Fellowship and by a DOE OJI award.

\appendix

\section{From elliptic functions to Theta functions}
\label{AAA}
Atiyah and Hitchin linearize \eqref{dw} by introducing
a solution of the auxiliary equation \cite{Atiyah:1985dv},
\begin{equation}
\label{du}
\frac{d^2 u}{d\theta^2}+\frac1{4\sin^2\theta} u =0
\end{equation}
where $dt/d\theta=1/u^2$. Then $w_{1,2,3}$ are given by
\begin{eqnarray}
\label{w123}
w_1&=&-u \frac{du}{d\theta}-\frac1{2\sin\theta} u^2 \nn\\
w_2&=&-u \frac{du}{d\theta}+\frac1{2\tan\theta} u^2 \\
w_3&=&-u \frac{du}{d\theta}+\frac1{2\sin\theta} u^2 \nn
\end{eqnarray}
The solution of \eqref{du} satisfying the appropriate limits is given
by a complete elliptic integral of the first kind,
\begin{equation}
\label{uk}
u=\sqrt{2\sin\theta} ~K\left( \sin\frac{\theta}{2} \right)
\end{equation}
where
\begin{equation}
\label{ellk}
K(k)=\int_0^{\pi/2}\frac{d\phi}{\sqrt{1-k^2\sin^2\phi}}
\end{equation}
The region of infinite separation corresponds to $k\to \infty$.

In order to see the equivalence with our solution \eqref{ww123}, recall
that the elliptic integral \eqref{ellk} is naturally associated to
an elliptic curve with complex modulus $\tau=i K/K'$, where
$K'=K((1-k^2)^{1/2})=K(k')$ with $k'=\cos(\theta/2)$. The relation between
$\tau$ and $K,K'$ can be rewritten as
\begin{equation}
k=\frac{\th_2^2}{\th_3^2}\ ,\quad
k'=\frac{\th_4^2}{\th_3^2}\ ,\quad
K=\frac{\pi}{2} \th_3^2
\end{equation}
Expressing the derivative of the elliptic integral $dK/dk=E/(kk^{'2})-K/k$
in terms of the elliptic integral of the second kind $E(k)$, related
to Jacobi Theta functions by
\begin{equation}
E=\frac{\th_3^4+\th_4^4}{3\th_3^4}K+\frac{\pi^2}{12K} E_2(\tau)
=K+\frac{\pi^2}{12K}(E_2+e_4)
\end{equation}
where we follow the conventions of \cite{Kiritsis} for modular forms.
we find that the solutions \eqref{w123} can be rewritten as
\begin{eqnarray}
\label{www123}
w_1&=&-\frac{\pi}{6}\left(E_2+\th_3^4+\th_4^4\right)\nn\\
w_2&=&-\frac{\pi}{6}\left(E_2+\th_2^4-\th_4^4\right)\\
w_3&=&-\frac{\pi}{6}\left(E_2-\th_2^4-\th_3^4\right)\nn
\end{eqnarray}
which is precisely the same as in \eqref{ww123}. The line element
$d\theta$ can be simply related to $dt$ as
$d\theta=-\pi\th_2^2\th_4^2 dt$, which agrees with the previous
relation $dt/d\theta=1/u^2$. It is also useful to introduce the
modular forms
\begin{equation}
\label{ei}
e_2=\th_3^4+\th_4^4\ ,\quad
e_3=\th_2^4-\th_4^4\ ,\quad
e_4=-\th_2^4-\th_3^4
\end{equation}
which are the roots of the Weierstrass polynomial
$x^3-3E_4~ x -2 E_6$ and are permuted under modular transformations.

\section{Riemann tensor of the Bianchi IX ansatz}
\label{BBB}
Let us now compute the curvature of the Atiyah-Hitchin
metric, in the orthonormal basis $e_0=abc~dt,e_i=a_i\sigma_i$. The
Levi-Civita connection is easily found to be
\begin{gather}
\omega_{i0}=a_i'/(abc)~ \sigma_i\ ,\quad
\omega_{jk}=\epsilon_{ijk} (a_k'/abc+1) ~\sigma_k
\end{gather}
and we can compute the curvature through $R=d\omega+[\omega,\omega]$,
\begin{equation}
R_{10}=\frac{d}{dt} \left(\frac{a'}{abc} \right) dt\wedge \sigma_1
+\frac{2 a b c a'-(a^2+b^2-c^2)b'c-(a^2+c^2-b^2)bc'}{2a^2 b^2 c^2}
\sigma_2\wedge \sigma_3\nn
\end{equation}
\begin{eqnarray}
R_{23}&=&\frac{d}{dt}\left( \frac{b^2+c^2-a^2}{2bc} \right) dt\wedge \sigma_1\\
&&-\frac{4 b' c'+bc(a^2+b^2-c^2)(a^2+c^2-b^2)-2a^2bc(b^2+c^2-a^2)}{4(abc)^2}
\sigma_2\wedge \sigma_3 \nn
\end{eqnarray}
In particular, a necessary condition for the metric to be self-dual is
\begin{equation}
\frac{a'}{abc}=  \frac{b^2+c^2-a^2}{2bc} - \lambda
\end{equation}
which agrees with \eqref{sdu} for $\lambda=1$. Defining $\hat a=
a'/(abc)$, we can rewrite this as
\begin{eqnarray}
R_{10} &=&R_{23}=
\hat a' dt \wedge \sigma_1 + (\hat a-\hat b-\hat c-2\hat b\hat c)
~\sigma_2\wedge \sigma_3\nn\\
R_{20} &=&R_{31}=
\hat b' dt \wedge \sigma_2 + (\hat b-\hat c-\hat a-2\hat c\hat a)
~\sigma_3\wedge \sigma_1\\
R_{30} &=&R_{12}=
\hat c' dt \wedge \sigma_3 + (\hat c-\hat a-\hat b-2\hat a\hat b)
~\sigma_1\wedge \sigma_2\nn
\end{eqnarray}
Using the identity
\begin{equation}
\frac{\hat a'}{a\cdot abc}=\frac{\hat a-\hat b-\hat c-2\hat b\hat c}{bc}
\end{equation}
we see that the Riemann tensor has the correct symmetry property.

\end{document}